\documentclass{article}
\usepackage{amsmath}

\input{tcilatex}

\begin{document}

\title{Dipolar radiation from spinning dust grains coupled to an
electromagnetic\ wave}
\author{A. Guerreiro$^{1}$, M. Eloy$^{2}$, J.T. Mendon\c{c}a$^{3}$, R.
Bingham$^{4}$ \\
$^{1}$Physics Department, Faculdade de Ci\^{e}ncias da Universidade do Porto%
\\
Rua do Campo Alegre, 687, 4169-007 Porto, Portugal\\
$^{2}$Faculdade de Engenharia da Universidade Cat\'{o}lica Portuguesa\\
Estrada Oct\'{a}vio Pato, 2635-631 Rio de Mouro, Portugal\\
$^{3}$Instituto Superior T\'{e}cnico\\
Av. Rovisco Pais, 1000 Lisboa, Portugal\\
$^{4}$Rutherford Appleton Laboratory, UK}
\maketitle

\begin{abstract}
In this letter we investigate how the complex rotation and quivering motion
of an elongated polarized dust grain in the presence of a monochromatic
electromagnetic wave can originate dipolar emission with two distinct
spectral components.

We present a model for the emission of radiation by elongated polarized dust
grains under the influence of both an external electromagnetic wave and a
constant background magnetic field. The dust, exhibiting rotational motion
at the external electromagnetic field frequency $\omega _{0}$ as well as
quivering motion at frequency $\Omega _{0}$, proportional to the em field
amplitude, will radiate with frequencies that will depend on the external
field wavelength and amplitude.

The radiated spectra exibits a frequency around $\omega _{0}$, and sidebands
at $\omega _{0}\pm \Omega _{0}$ and $\omega _{0}\pm 2\Omega _{0}$. Since the
amplitude and the frequency of the background electromagnetic field are
independent parameters, this model establishes a correlation between
different spectral components of galactic dipolar emission, which may help
to explain the correlation between a component of the Galactic microwave
emission and the $100\mu m$ thermal emission from interstellar dust, that
has been recently measured.
\end{abstract}

\section{Introduction}

Observations on cosmic background radiation have demonstrated the existence
of a correlation between a component of the Galactic microwave emission and
the $100\ \mu m$ thermal emission from interstellar dust, and several models
have been proposed to justify the results measured. Leitcht $et$ $al$. \cite%
{Leitcht97} suggested free-free emission from heated gas to be the cause for
this correlation but the model could not account for the expected $H_{\alpha
}$ emission unless the temperature would be in excess of $10^{6}K$, as
occurs in shock-heated gas from a supernova remnant. Drain and Lazarain
showed that for the free-free emission mechanism to be able to explain the
observed microwave excess it would require an energy input of at least two
orders of magnitude larger than that provided by supernovae and proposed an
alternative mechanism to explain the excess of microwave radiation observed,
based on electric dipole emission from rotating dust grains \cite{dl98},
which has been able to account for many aspects of the experimental
measurements.

More recently, data received from the Wilkinson Microwave Anisotropy Probe
(WMAP), has lead to the claim that the model proposed by Drain and Lazarain
could only account for 5\% of the Galactic microwave emission, and, once
again rekindled the debate about the origin of "anomalous" dust correlated
microwave emission \cite{Finkbeiner}. Also, Lazarain and Prunet \cite%
{Lazarian Prunet} have analyzed the importance of the thermal emission of
magnetized dust, locally aligned by the Galactic magnetic field in the
context of CMB contamination by Galactic dust emission, while Ponthieu \cite%
{Ponthieu} and others \cite{Baccigalupi} have stressed the importance of the
polarization effects by measuring a 3-5\% polarized dust signal on the
Galactic plane.

In parallel, several studies of levitation and dynamics of charged dust
grains have been carried out in both space and laboratory environments \cite%
{Nitter}, to investigate the motion of charged dust particles in low
temperature dusty plasmas discharges\cite{Chiang} and it has been observed
that in a dust plasma sheath dust grains levitate due to a balance between
gravitational and electrostatic forces, presenting a bouncing motion between
electrodes and a quivering motion across the electric field of the sheath.

Several processes have, since then, been proposed to explain dust dynamics,
with special emphasis on rotational excitation and damping of the dust
grains, which, in both astrophysical and laboratory plasmas, are known to
possess a rather elongated shape \cite{Spitzer}\cite{Harwit} exhibiting a
non-zero dipole moment. Examples include recoil from thermal collisions
and/or evaporation \cite{dl98b}, collisions with gas atoms or plasma and
plasma drag \cite{AndersonWatson}, absorption and emission of radiation \cite%
{Rouan92}\cite{Rouan97}, random H$_{2}$ formation \cite{dl98b}, \cite%
{HunterWatson} and systematic torques \cite{DrainWeingartner} \cite{dl98b}.
\ However, the influence of interstellar electromagnetic background
radiation on dust dynamics has not yet, to our knowledge, been fully
addressed.

Tskhaya and Shukla \cite{tskhakayashukla2001} have proposed a simple
analytical model to describe the dynamics of elongated dust grains in the
presence of circularly polarized electromagnetic waves and have concluded
that the grains not only rotate - or spin - with the frequency of
electromagnetic fields but also exhibits a complex quivering motion.
Radiative processes from such spinning dusts have not, however, been
included in the model.

This paper addresses the emission of radiation from a single rotating or
spinning dust grain under the influence of an external electromagnetic (em)
wave and a constant background magnetic field, correlating the external em
field amplitude as well as it wavelength with the dust-emitted frequency
spectra and predicting the radiated power due to both rotation and quivering
motions of the dust grain. The work presented is organized as follows:
starting from the same grain equations of motion as derived in \cite%
{tskhakayashukla2001} in Section 2, we address the emission of dipolar
radiation from spinning dust in Section 3, show the results of numerical
simulations of the emission spectrum of dust grains in section 4 and present
the conclusions in Section 5.

\section{Spinning dust dynamics}

We start by considering the propagation along the $z$ direction of a
circularly polarized electromagnetic (em) beam in a medium composed by
neutral elongated dust grains, such as a stellar dust cloud, in the presence
of a constant background magnetic field $\overrightarrow{B}=(0,0,B_{0})$.

The electric field can be written as:%
\begin{equation}
\overrightarrow{E}(\overrightarrow{r},t)=E_{0}\exp (i\overrightarrow{k}.%
\overrightarrow{r})\overrightarrow{e}(t)\text{,}  \label{field}
\end{equation}%
where $E_{0}$ is the electric field amplitude and $\overrightarrow{e}%
(t)=(\cos (\omega _{0}t),\sin (\omega _{0}t),0)$ is the polarization unit
vector.

We will assume that the dust magnetic moment of each grain is along the
direction of the background magnetic field ($z$) and neglect the precession
motion around this axis, meaning that the dust will rotate in the $xy$
plane. The dipole moment of the grain is expressed as $\overrightarrow{d}%
=d(\cos \phi ,\sin \phi ,0)$ with $\phi $ the orientation of the grain
relative to the $x$ direction.

The equation of motion describing the rotation of the grain will be \cite%
{tskhakayashukla2001}:%
\begin{equation}
I_{z}\frac{d^{2}\phi }{dt^{2}}=-dE_{0}\sin (\phi -\omega _{0}t)\text{,}
\end{equation}%
which can be written as: 
\begin{equation}
\frac{d^{2}\phi }{dt^{2}}=-\Omega _{0}^{2}\sin (\phi -\omega _{0}t)\text{,}
\end{equation}%
where $\Omega _{0}^{2}=dE_{0}/I_{z}$ and $I_{z}$ is the $z$ component of the
principal moment of inertia of the dust grain.

Under the influence of the electromagnetic wave $E_{0}$, the grain will
rotate with the same angular frequency, $\omega _{0}$ \cite%
{tskhakayashukla2001}, \cite{Sato2000}, \cite{Sato2001}. Besides this
spinning motion, the grain may also exhibit a quivering motion corresponding
to fluctuations on the mean spinning motion. It is therefore convenient to
write the angle of orientation of the grain as $\phi =\omega _{0}t+\delta
\phi $, which is equivalent to considering the quivering motion of the grain
in a reference frame which rotates with the polarization of the field $E_{0}$%
. The equation of motion then becomes:%
\begin{equation}
\overset{\cdot \cdot }{\delta \phi }=-\Omega _{0}^{2}\sin (\delta \phi )%
\text{.}  \label{eq Hill-1}
\end{equation}

As demonstrated in reference \cite{tskhakayashukla2001}, equation (\ref{eq
Hill-1}) can be integrated and, imposing the initial conditions:%
\begin{eqnarray}
\delta \phi (t &=&0)=\delta \phi _{0}\text{,} \\
\overset{\cdot }{\delta \phi }(t &=&0)=\overset{\cdot }{\delta \phi }_{0}%
\text{,}
\end{eqnarray}%
results in the following equation:%
\begin{equation}
\frac{1}{2\Omega _{0}^{2}}\left( \overset{\cdot }{\delta \phi }\right)
^{2}-\cos (\delta \phi )=\frac{1}{2\Omega _{0}^{2}}\left( \overset{\cdot }{%
\delta \phi _{0}}\right) ^{2}-\cos (\delta \phi _{0})\equiv \varepsilon
\end{equation}

The constant of integration $\varepsilon $ plays the role of an effective
energy and for values in the range $[$ $-1,1]$ the variation of $\delta \phi 
$ is bounded, resulting in a quivering motion of the grain with a frequency
roughly equal to $\Omega _{0}$ \cite{tskhakayashukla2001}.

In this paper we are interested in situations when $\varepsilon \simeq -1$
and the quivering motion of the grain is approximately harmonic:%
\begin{equation}
\delta \phi (t)\simeq \delta \phi _{0}\cos (\Omega _{0}t)+\frac{\overset{%
\cdot }{\delta \phi }_{0}}{\Omega _{0}}\sin (\Omega _{0}t)
\end{equation}

In the original frame of reference, the dust grains rotate with an angular
velocity given by $\omega =\omega _{0}+\overset{\cdot }{\delta \phi }$, thus
exhibiting a rotation motion with the frequency of the external em field $%
\omega _{0}$, and a quivering motion across the direction of the electric
field with a frequency $\Omega _{0}$, proportional to the external em field
amplitude. The quivering motion can be identified as the fluctuations of the
mean rotation motion.

\section{Radiation from spinning dusts}

As the grain rotates about the $z$ axis, the grain emits electromagnetic
radiation according to the following equation \cite{Jackson}:%
\begin{equation}
\overrightarrow{E}_{rad}(\overrightarrow{r},t)=\dint d^{3}r\frac{\rho (%
\overrightarrow{r})}{c^{2}\left\vert \overrightarrow{R}+\overrightarrow{r}%
\right\vert }\left[ \overrightarrow{n}\times \left[ \overrightarrow{n}\times 
\overrightarrow{a}\right] \right] \text{,}  \label{Radiated field}
\end{equation}%
where the retardation effects have been neglected, $\overrightarrow{r}$ is
the position vector of each element of volume of the grain relative to the
center of mass of the grain, $\overrightarrow{R}$ is the position vector of
the center of mass of the grain relative to the observer, $\overrightarrow{n}
$ is the direction of observation, $\rho (\overrightarrow{r})$ is the charge
density and $\overrightarrow{a}$ is the acceleration of of each element of
volume of the grain.

Assuming that the center of mass of the grain remains still the velocity of
each element of volume of the grain is:%
\begin{equation}
\overrightarrow{v}=\overrightarrow{\omega }\times \overrightarrow{r}\text{,}
\end{equation}%
with $\overrightarrow{\omega }=(0,0,\omega _{0}+\overset{\cdot }{\delta \phi 
})$ the angular velocity vector of the grain, hence the corresponding
acceleration is:%
\begin{equation}
\overrightarrow{a}=\overrightarrow{\delta \alpha }\times \overrightarrow{r}+%
\overrightarrow{\omega }\times \left[ \overrightarrow{\omega }\times 
\overrightarrow{r}\right] \text{,}  \label{angular acceleration}
\end{equation}%
where $\overrightarrow{\delta \alpha }=(0,0,\overset{\cdot \cdot }{\delta
\phi })$. The first term in equation (\ref{angular acceleration})
corresponds to a tangential acceleration $\overrightarrow{a_{t}}$, whereas
the second term describes a radial or centripetal acceleration $%
\overrightarrow{a_{r}}$, which are mutual orthogonal.

Using simple textbook algebra, we can derive the following equality: 
\begin{equation*}
\left[ \overrightarrow{n}\times \left[ \overrightarrow{n}\times 
\overrightarrow{a}\right] \right] =\overrightarrow{a_{t}}-(\overrightarrow{%
a_{t}}.\overrightarrow{n})\overrightarrow{n}+\overrightarrow{a_{r}}-(%
\overrightarrow{a_{r}}.\overrightarrow{n})\overrightarrow{n}.
\end{equation*}%
Integrating equation(\ref{Radiated field}) over the volume of the grain and
using the fact that $R>>r$, we obtain:%
\begin{equation}
\overrightarrow{E}_{rad}(\overrightarrow{r},t)\simeq \frac{d}{%
c^{2}\left\vert \overrightarrow{R}\right\vert }\left[ \overset{\cdot \cdot }{%
\delta \phi (t)}\overrightarrow{O}_{\overrightarrow{n}}[\overrightarrow{u_{t}%
}]+\omega ^{2}(t)\overrightarrow{O}_{\overrightarrow{n}}[\overrightarrow{%
u_{r}}]\right] ,  \label{Radiated field 2}
\end{equation}%
where we have used the following definitions:%
\begin{eqnarray}
\overrightarrow{u_{t}} &\equiv &(\cos (\phi ),\sin (\phi ),0)\text{,} \\
\overrightarrow{u_{r}} &\equiv &(-\sin (\phi ),\cos (\phi ),0)\text{,} \\
\overrightarrow{O}_{\overrightarrow{n}}[\overrightarrow{x}] &\equiv &%
\overrightarrow{x}-(\overrightarrow{x}.\overrightarrow{n})\overrightarrow{n}%
\text{,}
\end{eqnarray}%
and where $\overrightarrow{d}=\dint d^{3}r\rho (\overrightarrow{r})%
\overrightarrow{r}$ is the dipole moment of the grain.

Using the fact that $\omega ^{2}=\omega _{0}^{2}+2\omega _{0}\overset{\cdot }%
{\delta \phi }+\overset{\cdot }{\delta \phi }^{2}$, yields:%
\begin{equation}
\overrightarrow{E}_{rad}(\overrightarrow{r},t)\simeq \frac{d}{%
c^{2}\left\vert \overrightarrow{R}\right\vert }\left[ \overset{\cdot \cdot }{%
\delta \phi (t)}\overrightarrow{O}_{\overrightarrow{n}}[\overrightarrow{u_{t}%
}]+\left[ \omega _{0}^{2}+2\omega _{0}\overset{\cdot }{\delta \phi (t)}+%
\overset{\cdot }{\delta \phi (t)}^{2}\right] \overrightarrow{O}_{%
\overrightarrow{n}}[\overrightarrow{u_{r}}]\right] \text{.}
\end{equation}

Since the CMB has a low intensity we will investigate the case of weak field
amplitude, corresponding to a small quivering frequencies and $\Omega
_{0}<<\omega _{0}$, then the dust-radiated electric field can be separated
into three sets of spectral lines around $\omega _{0}$. The first
corresponds to the emission of the standard dipolar radiation with frequency 
$\omega _{0}$ (remember that $\overrightarrow{u_{t}}$ and $\overrightarrow{%
u_{r}}$ rotate approximately with frequency $\omega _{0}$) associated with
the rotational motion of the grain:%
\begin{equation}
\overrightarrow{E}_{rad,1}(\overrightarrow{r},t)\simeq \frac{d}{%
c^{2}\left\vert \overrightarrow{R}\right\vert }\omega _{0}^{2}%
\overrightarrow{O}_{\overrightarrow{n}}[\overrightarrow{u_{r}}]\propto \frac{%
d}{c^{2}\left\vert \overrightarrow{R}\right\vert }\omega _{0}^{2}e^{i\omega
_{0}t}\text{.}
\end{equation}

The second set of spectral lines is associated with the quivering motion,
resulting in a spectral broadening of the standard dipole radiation with the
generation of two sidebands with frequencies $\omega _{0}+\Omega _{0}$
(anti-Stokes) and $\omega _{0}-\Omega _{0}$ (Stokes):%
\begin{eqnarray}
\overrightarrow{E}_{rad,2}(\overrightarrow{r},t) &\simeq &\frac{d}{%
c^{2}\left\vert \overrightarrow{R}\right\vert }\left[ \overset{\cdot \cdot }{%
\delta \phi (t)}\overrightarrow{O}_{\overrightarrow{n}}[\overrightarrow{u_{t}%
}]+2\omega _{0}\overset{\cdot }{\delta \phi (t)}\overrightarrow{O}_{%
\overrightarrow{n}}[\overrightarrow{u_{r}}]\right] \\
&\propto &\frac{d}{c^{2}\left\vert \overrightarrow{R}\right\vert }\left[
\Omega _{0}^{2}+2\omega _{0}\Omega _{0}\right] \arg \cos (\varepsilon
/2)e^{i(\omega _{0}\pm \Omega _{0})t}\text{.}
\end{eqnarray}

Finally, the third set of spectral lines is also associated with the
quivering motion, but the spectral broadening produces essentially two
sidebands with frequencies $\omega _{0}+2\Omega _{0}$ (double anti-Stokes)
and $\omega _{0}-2\Omega _{0}$ (double Stokes):%
\begin{eqnarray}
\overrightarrow{E}_{rad,3}(\overrightarrow{r},t) &\simeq &\frac{d}{%
c^{2}\left\vert \overrightarrow{R}\right\vert }\left[ \overset{\cdot }{%
\delta \phi (t)}^{2}\overrightarrow{O}_{\overrightarrow{n}}[\overrightarrow{%
u_{r}}]\right]  \\
&\propto &\frac{d}{c^{2}\left\vert \overrightarrow{R}\right\vert }\left[
\Omega _{0}^{2}\right] \arg \cos ^{2}(\varepsilon /2)e^{i(\omega _{0}\pm
2\Omega _{0})t}\text{.}
\end{eqnarray}

The instantaneous energy flux is given by the Poyting vector:%
\begin{equation}
\overrightarrow{S}=\frac{c}{4\pi }\overrightarrow{E}_{rad}\times 
\overrightarrow{B}_{rad}=\frac{c}{4\pi }\left\vert E_{rad}\right\vert ^{2}%
\overrightarrow{n}\text{,}
\end{equation}%
and the power radiated per unit solid angle can be written as:%
\begin{eqnarray}
\frac{dP}{d\Omega } &=&\frac{c}{4\pi }\left\vert \overrightarrow{R}%
\right\vert ^{2}\left\vert E_{rad}\right\vert ^{2}  \notag \\
&=&\frac{d^{2}}{4\pi c^{3}}\left\{ \left\vert \overset{\cdot \cdot }{\delta
\phi (t)}\right\vert ^{2}\sin ^{2}(\Theta _{t})+\left\vert \omega
_{0}^{2}+2\omega _{0}\overset{\cdot }{\delta \phi (t)}\overset{\cdot }{%
+\delta \phi (t)}^{2}\right\vert ^{2}\sin ^{2}(\Theta _{r})\right. -  \notag
\\
&&-\left. 2\overset{\cdot \cdot }{\delta \phi }\left[ \omega
_{0}^{2}+2\omega _{0}\overset{\cdot }{\delta \phi (t)}+\overset{\cdot }{%
\delta \phi (t)}^{2}\right] \cos (\Theta _{t})\cos (\Theta _{r})\right\} ,
\end{eqnarray}%
where $\Theta _{t}$ and $\Theta _{r}$ are respectively the angle between $%
\overrightarrow{u_{t}}$ and $\overrightarrow{u_{r}}$, and the direction of
observation $\overrightarrow{n}$.

By separating the radiation into the two spectral components we recover the
standard result for the power radiated per unit solid angle emitted by a
dipole rotating with constant angular velocity $\omega _{0}$:%
\begin{equation}
\frac{dP_{1}}{d\Omega }=\frac{d^{2}}{4\pi c^{3}}\omega _{0}^{4}\sin
^{2}(\Theta _{r})
\end{equation}%
and we identify the power radiated per unit solid angle due to the quivering:

\begin{eqnarray}
\frac{dP_{2}}{d\Omega } &=&\frac{d^{2}}{4\pi c^{3}}\left\{ \left\vert 
\overset{\cdot \cdot }{\delta \phi }\right\vert ^{2}\sin ^{2}(\Theta
_{t})+\left\vert 2\omega _{0}\overset{\cdot }{\delta \phi (t)}\right\vert
^{2}\sin ^{2}(\Theta _{r})\right. -  \notag \\
&&-\left. 4\omega _{0}\overset{\cdot \cdot }{\delta \phi }\overset{\cdot }{%
\delta \phi (t)}\cos (\Theta _{t})\cos (\Theta _{r})\right\}
\end{eqnarray}%
and 
\begin{equation}
\frac{dP_{3}}{d\Omega }=\frac{d^{2}}{4\pi c^{3}}\left\vert \overset{\cdot }{%
\delta \phi (t)}\right\vert ^{4}\sin ^{2}(\Theta _{r})\text{.}
\end{equation}

Since the quivering motion is much slower than the rotation of the grain,
the average radiated power per unit solid angle over a period of rotation of
the grain yields:%
\begin{eqnarray}
\frac{d\left\langle P_{1}\right\rangle }{d\Omega } &\approx &\frac{d^{2}}{%
4\pi c^{3}}\omega _{0}^{4}\left[ 1-\cos ^{2}(\Phi )\right] , \\
\frac{d\left\langle P_{2}\right\rangle }{d\Omega } &\approx &\frac{d^{2}}{%
4\pi c^{3}}\left[ \Omega _{0}^{4}+4\omega _{0}^{2}\Omega _{0}^{2}\right] %
\left[ 1-\cos ^{2}(\Phi )\right] , \\
\frac{d\left\langle P_{3}\right\rangle }{d\Omega } &\approx &\frac{d^{2}}{%
4\pi c^{3}}\Omega _{0}^{4}\left[ 1-\cos ^{2}(\Phi )\right] ,
\end{eqnarray}%
where the theorem of equipartion of energy, $\left\langle \delta \phi
(t)\right\rangle =\arg \cos (\varepsilon /2)$, has been used and $\Phi $
representing the angle between the direction of observation $\overrightarrow{%
n}$ and the plane of rotation of the grain, or the direction of the
background magnetic field.

The angular emission distribution is represented in Figure 1, showing that
the emission is larger along the direction of the background magnetic field
and null along the plane of rotation of the grain, which indicates that each
dust grain behaves as a small probe sensing the local electromagnetic
environment - both the frequency and intensity of the electromagnetic field
as well as the direction of the background magnetic field - and imprinting
them in the emitted dipolar radiation.

\FRAME{ftbpFU}{3.1537in}{3.1537in}{0pt}{\Qcb{Scheme of the mean angular
emission of a rotating dust graing: the emission is larger along the
direction of the background magnetic field and null along the plane of
rotation of the grain.}}{}{diapositivo2.jpg}{\special{language "Scientific
Word";type "GRAPHIC";maintain-aspect-ratio TRUE;display "USEDEF";valid_file
"F";width 3.1537in;height 3.1537in;depth 0pt;original-width
6.7463in;original-height 6.7463in;cropleft "0";croptop "1";cropright
"1";cropbottom "0";filename '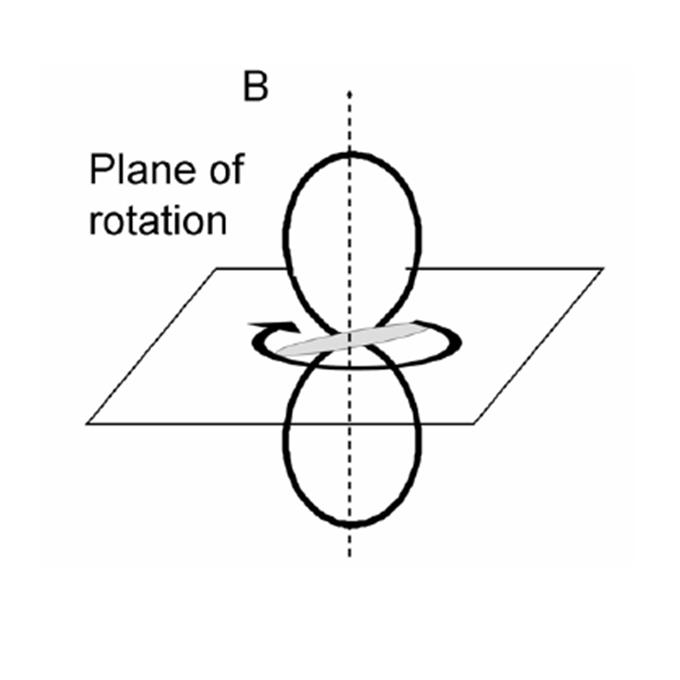';file-properties "XNPEU";}}

\section{Simulation}

The analysis of the dynamics of a single dust grain presented in the
previous section has provided much insight about the dipolar emission
spectrum in a dust cloud, enabling the identification of two regimes (for $%
\Omega _{0}>>\omega _{0}$ and $\Omega _{0}<<\omega _{0}$), as well as the
dominant frequencies of the emitted spectrum. A more real model can devised
by averaging the emission spectrum over an ensemble of identical grains with
the quivering temperature $T$:%
\begin{equation}
I(\omega )=\underset{[-1,1]}{\tint }d\varepsilon \ n(\varepsilon )\
\left\langle I(\omega ,\varepsilon )\right\rangle \text{,}  \label{eq. mean}
\end{equation}%
where $\left\langle I(\omega ,\varepsilon )\right\rangle $ is the Gibbs
average of the emission spectrum over the ensemble, $n(\varepsilon )\simeq
n_{0}\exp \left[ -(\varepsilon +1)I\Omega _{0}^{2}/2K_{B}T\right] $ the
Boltzmann distribution for the probabilities of occupancy of the rotational
energies in the large angular momentum limit \cite{Beiser} (notice that the
quivering angular momentum of the grain is $\varepsilon I\Omega _{0}$) and $%
K_{B}$ the Boltzmann constant. Since the grains are quasi-periodic systems,
the Gibbs average $\left\langle I(\omega ,\varepsilon )\right\rangle $ in
equation (\ref{eq. mean}) can be replaced by a time sampling according to
the Ergodic hypothesis, which states that the evolution of a complex
classical dynamical system takes it, with equal probability, through all
states which are accessible from the starting point subject to the constrain
of energy conservation \cite{Binney}\cite{Metropolis}.

The emission spectrum of each dust grain has been simulated using a simple
fourth-order Runge-Kutta integrator to numerically solve the following
equation:%
\begin{equation}
\frac{d^{2}\delta \phi }{d\tau ^{2}}=-\sin (\delta \phi ),
\end{equation}%
with $\tau =\Omega _{0}t$ is the natural time scale of the process. To
sample over different energy configurations, a method based on the
Metropolis algorithm \cite{Metropolis}, has been used, in which a random
configuration of initial conditions is generated and the effective energy $%
\varepsilon $ is calculated. Should the energy change relative to the
previous configuration $\varepsilon ^{\prime }$ be negative, the new
configuration is automatically accepted, else, the new configuration is
accepted with probability $\exp \left[ (\varepsilon -\varepsilon ^{\prime
})I\Omega _{0}^{2}/2K_{B}T\right] $.

Simulation results are shown in Figure 2 for both regimes, in which the
quivering motion is assumed to be in equilibrium with the background at
temperature $T=3K$. Also, a cutoff of the effective energy at $\varepsilon
=0.5$ has been imposed, in order to maintain the results close to the domain
of validity of the model. The emission pertaining to each single grain is
shown in grey (which can be interpreted as the sampling fluctuations) and
the mean radiated spectrum in black. In both regimes the thermal sampling
leads to the expected broadening of all the emission spectral picks, though
it is still possible to identify the predicted features.

Since all nonlinear terms of equation (\ref{eq Hill-1}) have been included
in the simulations, it is possible to detect the signature, though weak, of
higher order sidebands in the fluctuation spectrum (grey lines) at $3\omega
_{0}$, $4\omega _{0}$... for the low-intensity regime, corresponding to
nonlinear wave mixing of the external em wave through the coupling with the
grain.

\FRAME{ftbpFU}{3.1537in}{3.1537in}{0pt}{\Qcb{Emission spectrum of the grain:
The quivering frequency $\Omega _{0}$ is much smaller than the rotation
frequency $\protect\omega _{0}$. Sidebands at $\protect\omega _{0}\pm \Omega
_{0}$ and $\protect\omega _{0}\pm 2\Omega _{0}$ are clearly identifiable. }}{%
\Qlb{Figure2}}{diapositivo1.jpg}{\special{language "Scientific Word";type
"GRAPHIC";maintain-aspect-ratio TRUE;display "USEDEF";valid_file "F";width
3.1537in;height 3.1537in;depth 0pt;original-width 6.7463in;original-height
6.7463in;cropleft "0";croptop "1";cropright "1";cropbottom "0";filename
'Diapositivo1.JPG';file-properties "XNPEU";}}

\section{Conclusions}

In summary, we have presented a model for the emission of radiation by
elongated polarized dust grains under the influence of both an external
electromagnetic (em) wave and a constant background magnetic field. The
emission spectrum depends on the wavelength of the external em wave but -
and most importantly - also on the amplitude, or intensity, of the external
em field, which is, in fact, the parameter that determines the frequencies
of much of the components of the emitted radiation.

The emission spectrum of the grain exhibits a spectral line centered at the
external em field wavelength, or frequency, $\omega _{0}$, and Raman-like
sidebands at $\omega _{0}\pm \Omega _{0}$ and $\omega _{0}\pm 2\Omega _{0}$.
We have also identified that the optimum direction of emission to be
parallel to the background magnetic field, rendering dust grains
exceptionally useful for local probing of magnetic fields near far
astrophysical objects.

Numerical simulations indicate that, as expected, thermal averaging of the
spectrum, and the inclusion of the nonlinear dust dynamics can lead to the
broadening and smoothing of the emitted spectrum.

Since the amplitude and the frequency of the background em field are
independent parameters, this model allows to predict a correlation between
different spectral components of galactic dipolar emission, which may
explain the correlation between a component of the Galactic microwave
emission and the 100 $\mu m$ thermal emission from interstellar dust.

\textbf{Acknowledgement}

The authors would like to thank Dr. Pedro Carvalho for his insightful
comments on Cosmic Microwave Background and Dr. Jo\~{a}o Lopes dos Santos
for his remarks on the Metropolis algorithm.

\end{document}